\documentclass{xjkas}

\def\year{2014} 
\def\month{September} 
\def\volume{00} 
\def\beginpage{1} 
\def\endpage{5} 
\def\received{September 5, 2014} 
\def\accepted{October 10, 2014} 
\date{Received \received ; accepted \accepted}

\usepackage{flushend} 
\usepackage[breaklinks,colorlinks,citecolor=blue,linkcolor=blue,menucolor=blue,urlcolor=blue]{hyperref}

\def\vimap{{\sf VIMAP}}

\title{
VIMAP: an Interactive Program Providing Radio\\ Spectral Index Maps of Active Galactic Nuclei
}

\author[]{Jae-Young Kim}
\author[]{Sascha Trippe}

\affil[]{Department of Physics and Astronomy, Seoul National University, 599 Gwanak-ro, Gwanak-gu, Seoul 151-742, Korea; \email{jykim@astro.snu.ac.kr; trippe@astro.snu.ac.kr}}

\def\corrauthor{
S. Trippe
}

\def\runningauthor{
Kim \& Trippe
}

\def\runningtitle{
VIMAP: Program Providing Radio Spectral Index Maps of AGN
}

\def\keywords{
Galaxies: active --- Radio continuum: galaxies --- Methods: data analysis; numerical --- Techniques: interferometric
}

\def\abstracttext{
We present a GUI-based interactive Python program, \vimap, which generates radio spectral index maps of active galactic nuclei (AGN) from Very Long Baseline Interferometry (VLBI) maps obtained at different frequencies. \vimap\ is a handy tool for the spectral analysis of synchrotron emission from AGN jets, specifically of spectral index distributions, turn-over frequencies, and core-shifts. In general, the required accurate image alignment is difficult to achieve because of a loss of absolute spatial coordinate information during VLBI data reduction (self-calibration) and/or intrinsic variations of source structure as function of frequency. These issues are overcome by \vimap\ which in turn is based on the two-dimensional cross-correlation algorithm of \citet{croke2008}. In this paper, we briefly review the problem of aligning VLBI AGN maps, describe the workflow of \vimap, and present an analysis of archival VLBI maps of the active nucleus 3C~120.
}

\begin{document}
\jkashead 


\section{Introduction\label{sec:intro}}

Active galactic nuclei (AGN) emit strong radio continuum synchrotron radiation and show spatial structure at scales ranging from sub-parsecs to kiloparsecs. Typical features are a core and outflows, especially jets, of usually complex structure.
Multiple theoretical and observational studies indicate that the accretion of gas onto rotating supermassive black holes and acceleration of the accreted matter by magneto-hydrodynamical processes is responsible for launching relativistic outflows and powering the observed high-energy synchrotron radiation (see, e.g., \citealt{book2014} for a recent review).

Multi-frequency Very Long Baseline Interferometry (VLBI) observations are crucial for studying the physics and evolution of the outflows of AGN. Synchrotron radiation is characterized by its spectral index, $\alpha \equiv \log(S_{\nu_{2}}/S_{\nu_{1}})/\log(\nu_{2}/\nu_{1})$, where $S_{\nu}$ is the flux at frequency $\nu$ and $\nu_{1}$ and $\nu_{2}$ are two different observing frequencies with $\nu_{2}>\nu_{1}$. Measurements of the spatial distribution of $\alpha$ provide valuable physical information. \citet{walker2000} revealed the spatial geometry of ionized gas associated with the accretion disk of 3C~84 by analyzing its continuum absorption spectra. \citet{fromm2013} interpreted a radio flare in CTA~102 as interaction of a traveling shock with a stationary structure (potentially a recollimation shock). \citet{osullivan2009} measured particle densities and magnetic field strengths in the parsec-scale jets of six active galaxies in order to test the classical model of AGN jets proposed by \citet{bk79}. Analyses of the structural and spectral variability of radio jets complement single-dish studies of temporal AGN variability (e.g., \citealt{park2012,park2014,kim2013}) and of the interplay between accretion and jet formation (e.g., \citealt{allen2006,trippe2014}).

In practice, obtaining spatially resolved spectral index information for AGN outflows is challenging mainly for two reasons. Firstly, an AGN core corresponds to a radio photosphere, e.g., an optical depth $\tau_{\nu} =1$ surface, for which the observed position is a function of frequency (a phenomenon known as \textit{core-shift} effect; \citealt{lobanov1998a}). Secondly, phase self-calibration, which is an important step in VLBI data reduction, removes absolute coordinate information (\citealt{kameno2003,kadler2004}). Both effects imply the need for a careful spatial alignment of radio maps obtained at different frequencies.
Traditionally, two methods for aligning multi-frequency VLBI images have been used. The first one uses optically thin \textit{compact} components, which are assumed to not change their positions as function of frequency, as reference points in a map (e.g., \citealt{kadler2004}). 
The second method employs spatial correlations of optically thin \textit{extended} jet structure for achieving alignment and/or for measuring the core-shift (\citealt{walker2000,croke2008}).

Currently, standard radio astronomical software packages, especially {\sf AIPS}, {\sf CASA},\footnote{Maintained by the National Radio Astronomical Observatory of the USA.} and {\sf Difmap} \citep{shep1997} lack specialized tasks dedicated to image alignment. Motivated by this, we developed a highly interactive graphical user interface (GUI) based Python program, \vimap, which adopts the two-dimensional cross-correlation scheme of \citet{croke2008}. \vimap\ is open to the public and is capable of performing map alignment and the generation of spectral index maps in a straightforward manner. In Section \ref{sec:workflow} we outline how \vimap\ works. In Section \ref{sec:format} we present an analysis of archival VLBI maps of the AGN 3C~120 for demonstration.

\section{The VIMAP Workflow \label{sec:workflow}} 

\subsection{Prerequisites \label{sec:info}}

\vimap\ is written in Python version 2.7.3 and uses the following well-known numerical and astronomical add-on packages:
\begin{itemize}
  \item Numpy version $1.8.1$ or higher\footnote{\url{http://www.numpy.org}}
  \item Scipy version $0.14.0$ or higher\footnote{\url{http://www.scipy.org}}
  \item Matplotlib version $1.3.1$ or higher\footnote{\url{http://matplotlib.org}}
  \item PyFITS version $3.2.3$ or higher\footnote{\url{http://www.STScI.edu/institute/software_hardware/PyFITS}}
  \item wxPython version $2.8.12.1$ or higher\footnote{\url{http://www.wxPython.org}}
\end{itemize}
These packages are available from the corresponding web sites. \vimap\ itself, plus supplementary information material, is freely available on the Internet.\footnote{\url{http://www.astro.snu.ac.kr/~trippe/VIMAP}}

\begin{figure}[t!]
\centering
\includegraphics[angle=-90,width=80mm]{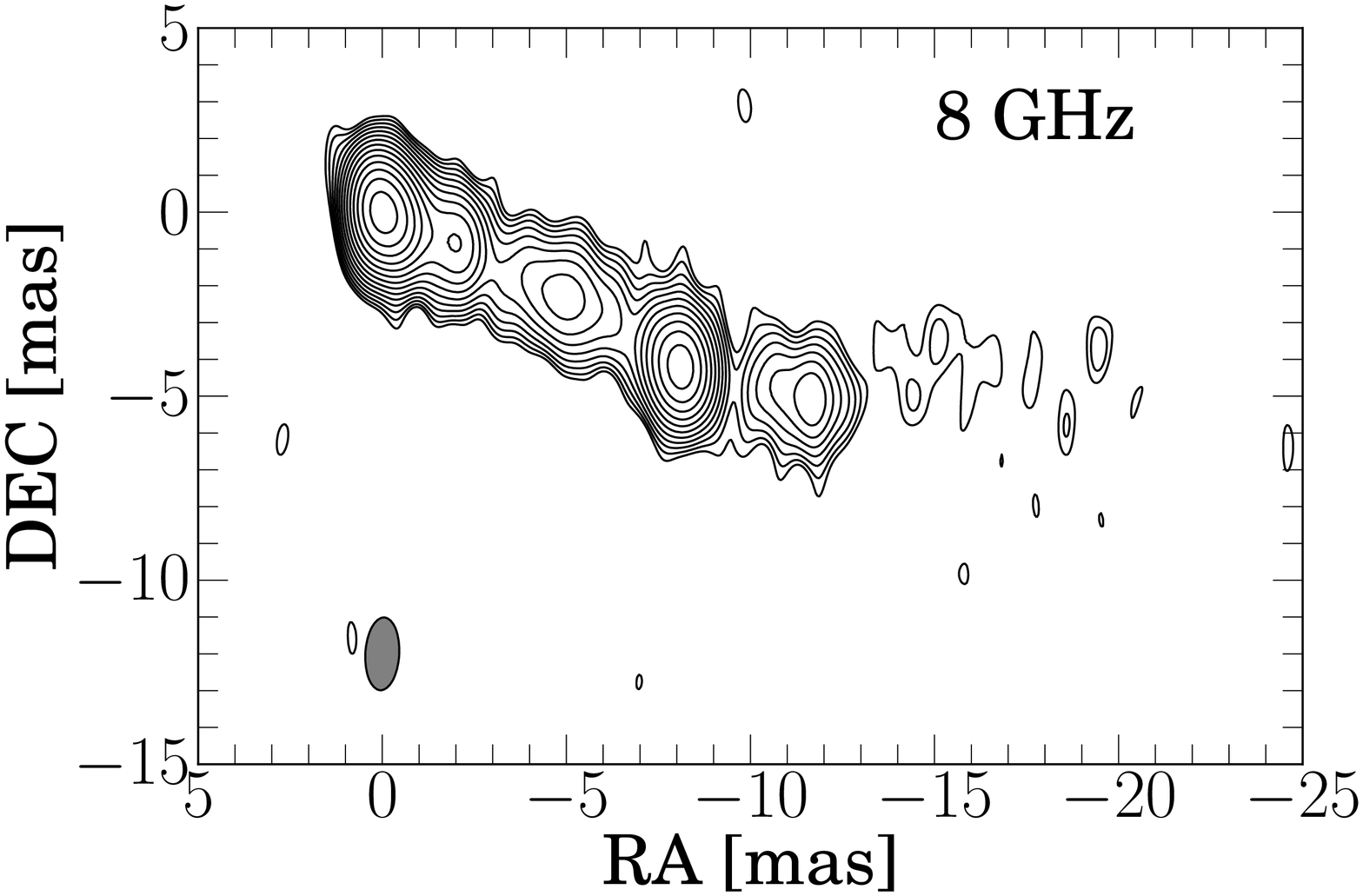} \\ 
\includegraphics[angle=-90,width=80mm]{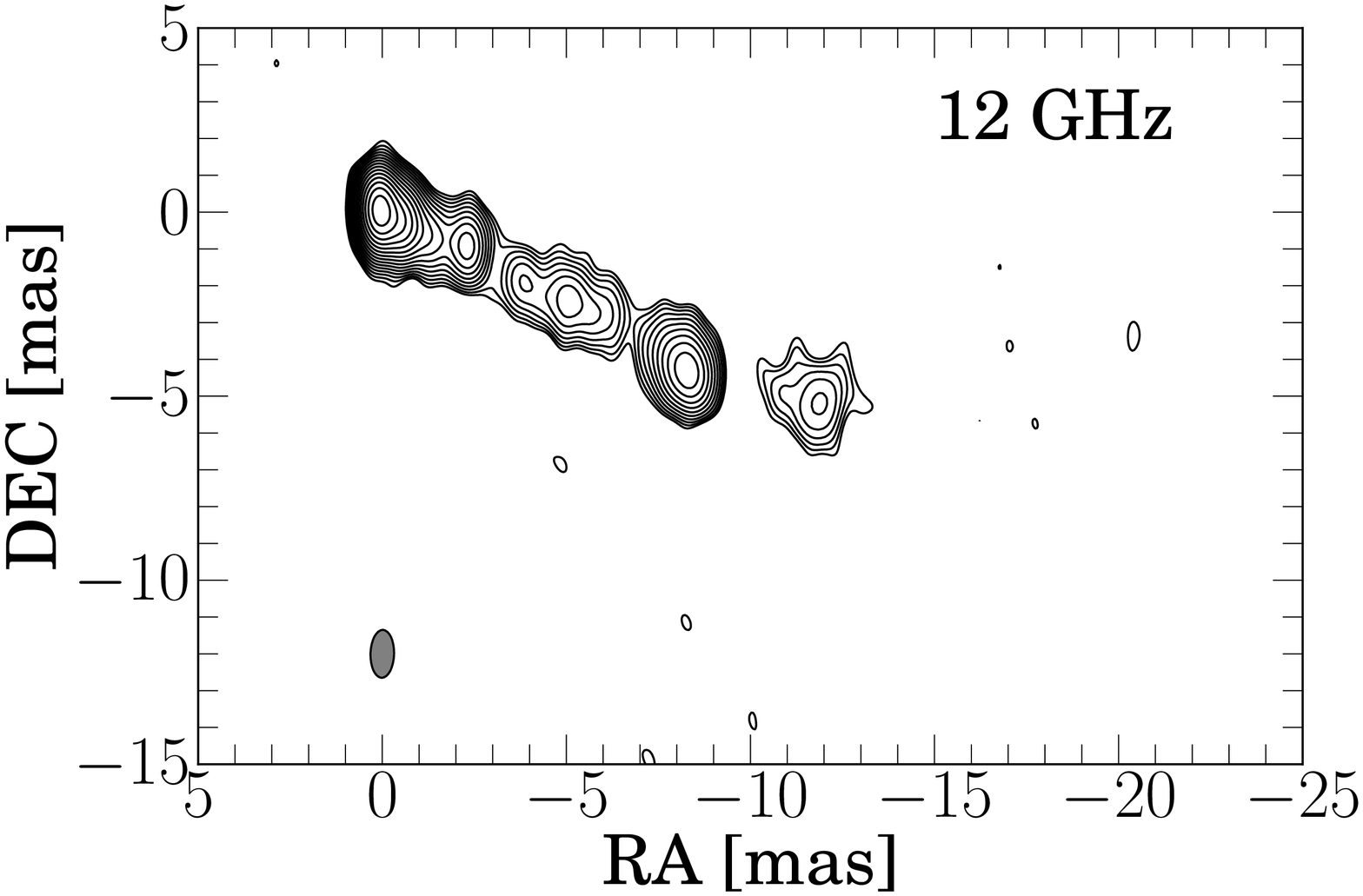} \\ 
\includegraphics[angle=-90,width=80mm]{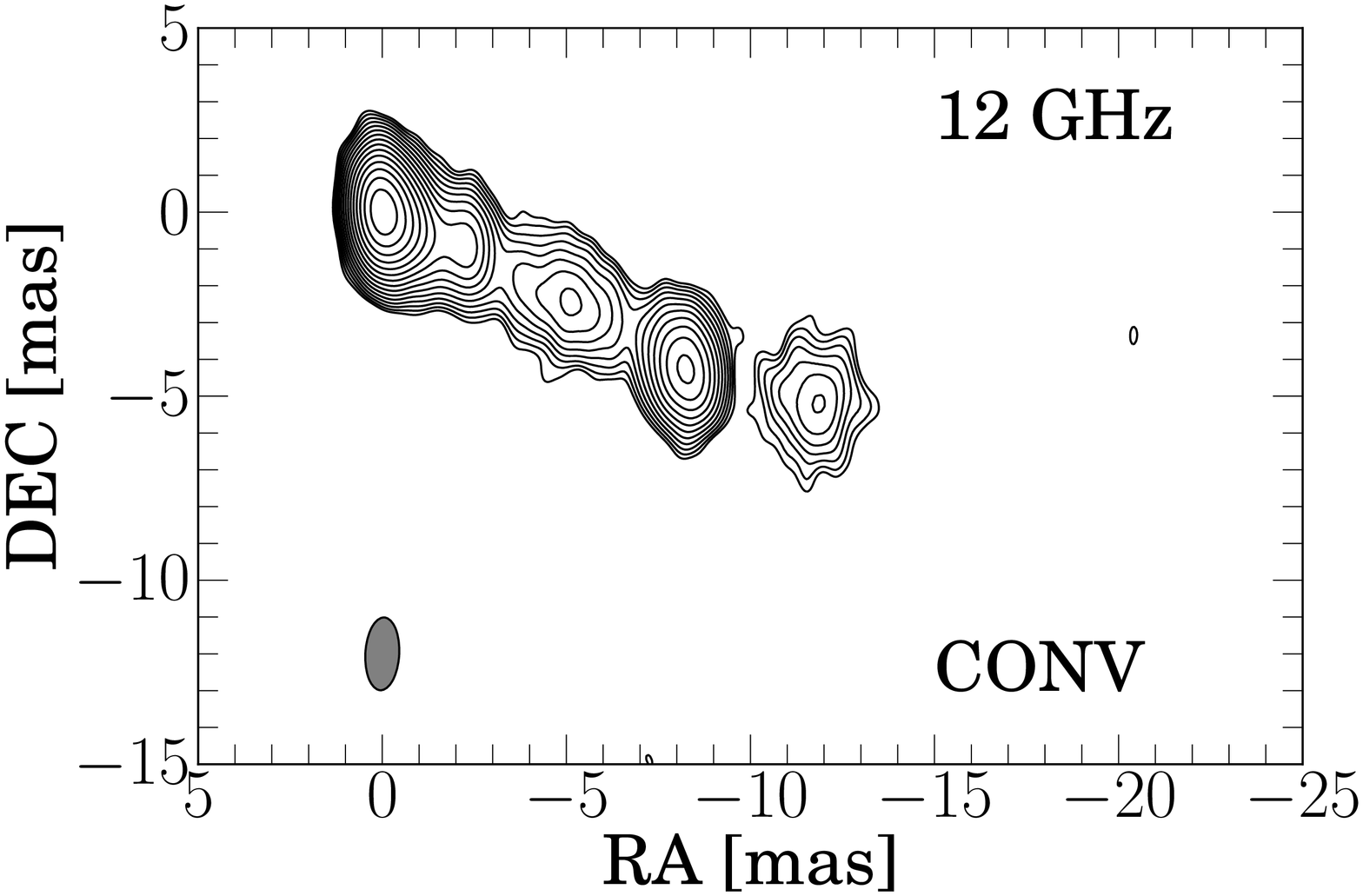} 
\caption{VLBI maps of the active galaxy 3C~120 at 8 and 12 GHz. Contour levels decrease by factors of $\sqrt{2}$ from the peak. Grey ellipses indicate beam sizes.
\textit{Top}: 8-GHz map. The peak intensity is 0.70 $\rm Jy/beam$ with the lowest contour corresponding to 0.9\% of the peak value.
\textit{Center}: 12-GHz map. The peak intensity is 0.96 $\rm Jy/beam$ with the lowest contour level corresponding to 0.6\% of the peak value.
\textit{Bottom}: Same as center panel, but using the restoring beam of the 8-GHz map. The peak intensity is then increased to 1.09 $\rm Jy/beam$.
}
\label{fig:illustration}
\end{figure}

\begin{figure}[t!]
\centering
\includegraphics[width=82mm]{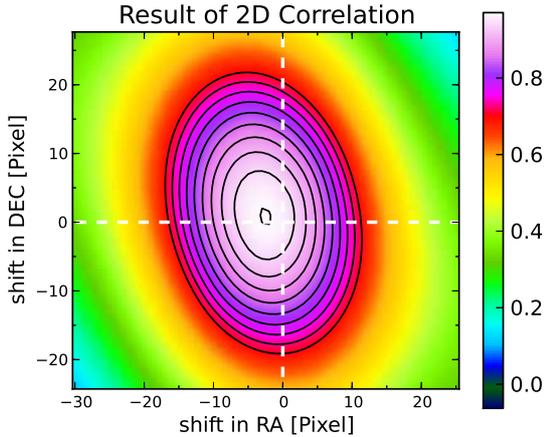}
\caption{
2D cross-correlation map; the color scales indicates the value of the correlation coefficient. Offsets are in units of pixels. The 12-GHz map is shifted relative to the 8-GHz map. Black contours correspond to correlation values of 0.7, 0.73, 0.76, 0.79, etc. 
}
\label{fig:result1}
\end{figure}

\subsection{Preparation of Radio Maps \label{sec:maps}}

In order to begin the analysis, the sizes (in pixels) and pixel scales (angular extension of a pixel) of the two maps need to be made equal.
Once this is achieved, the map obtained at frequency $\nu_{2}>\nu_{1}$ needs to be convolved with the restoring beam of the map obtained at the frequency $\nu_{1}$, or has to be reconstructed from its $\delta$ function CLEAN components using the beam of the map at $\nu_{1}$. This is easily achieved using the standard $\sf AIPS$ and/or $\sf Difmap$ data reduction packages.
The pixel scale should be much smaller than the synthesized beam size at $\nu_{1}$;
we recommend a size of 1/20 (or less) of the beam size.

\subsection{Masking of Cores \label{sec:mask}}

In general, AGN cores need to be excluded from image alignment calculations because their positions may vary as function of frequency.
In \vimap, users can place an elliptical mask onto a core with variable (i) center position, (ii) semimajor/minor axis lengths, and (iii) orientation. Depending on how well the core can be identified and separated from the outflows in a given map, it may be necessary to iterate the alignment several times with different choices of masks.

\subsection{Cross-Correlation \label{sec:correl}}

The relative shift between two images is estimated via a two-dimensional cross-correlation.
The definition of the correlation coefficient $r_{xy}$ implemented in \vimap\ is
\begin{equation}\small
r_{xy} = \frac{\sum_{i,j} (I_{\nu_{1},ij}-\overline{I}_{\nu_{1}})(I_{\nu_{2},ij} -\overline{I}_{\nu_{2}})
}{
\sqrt{\left[\sum_{i,j} (I_{\nu_{1},ij} -\overline{I}_{\nu_{1}})^{2}\right]\left[\sum_{i,j} (I_{\nu_{2},ij} -\overline{I}_{\nu_{2}})^{2}\right]}}
\label{cor_formula}
\end{equation}
\citep{croke2008} where
$I_{\nu_{1,2},ij}$ is the intensity at frequency $\nu_{1,2}$ and at spatial coordinate $(i,j)$, 
$\overline{I}_{\nu_{1,2}}$ is the mean intensity of a map at $\nu_{1,2}$.

The output of the cross-correlation analysis is a 2D map displaying the correlation coefficient $r_{xy}$ as function of relative offsets in right ascension (RA) and declination (DEC). The spatial shift between the images is given by the location of the maximum value of $r_{xy}$. If necessary, \vimap\ users can set offsets manually.

\subsection{Spectral Index Maps \label{sec:indexmap}}

Eventually, \vimap\ generates a spectral index map showing the index $\alpha$ (as defined in the introduction) as function of position, as well as a spectral index error map.

The spectral index error map is based on the assumption that the uncertainty \emph{of the flux density} in each of the two radio maps used is a combination of (i) a systematic error coming from imperfect visibility amplitude calibration and (ii) random thermal noise from the observed source, sky, and instruments.
This implies a total intensity error $\sigma_{\nu, ij}$, as function of frequency $\nu$ and image location $(i,j)$, of
  \begin{equation}
	\sigma_{\nu, ij} =  \delta_{\nu}\,I_{\nu, ij}+RMS_{\nu, ij} ~.
	 \label{erreq_eq1}
  \end{equation}
Here $\delta_{\nu}\lesssim0.1$ is the systematic amplitude calibration error on amplitude and
$RMS_{\nu, ij}$ is the root-mean-squared thermal noise. $RMS_{\nu, ij}$ can be measured in empty regions of an image. The factor $\delta_{\nu}$ may be known a priori for a given interferometer; if not, it can be estimated by measuring the flux from a compact flux calibrator and cross-comparison to results found by other observatories. 
Eventually, the uncertainty \emph{of the spectral index} $\alpha_{ij}$ -- i.e., the final spectral index error map $Err(\alpha_{\nu_{1,2},ij})$ -- 
is found via standard error propagation:
  \begin{equation}
	 Err(\alpha_{\nu_{1,2}, ij}) = \frac{1}{\log (\nu_{2}/\nu_{1}) } \times 
	 \left[ \frac{\sigma^{2}_{\nu_{1,ij}}}{I^{2}_{\nu_{1,ij}}} +  \frac{\sigma^{2}_{\nu_{2,ij}}}{ I^{2}_{\nu_{2,ij}}} \right]^{1/2} .
	 \label{error_eq2} 
  \end{equation}

When generating the final spectral index map, \vimap\ employs two cutoffs that exclude trivial or unphysical spectral index values: (i) a marginal intensity boundary, 
outside of which the source flux is assumed to be too weak for extracting spectral index information; and (ii) an upper limit on the error on $\alpha_{ij}$.

\section{Application \label{sec:format}}

In the following, we provide an analysis of a VLBI data set with \vimap. We used two Very Long Baseline Array (VLBA) maps of the AGN 3C 120 obtained in 2006 at 8.4 and 12.1 GHz, respectively. We obtained raw data from the archive of the MOJAVE program\footnote{\url{http://www.physics.purdue.edu/astro/MOJAVE/}} \citep{lister2009} and converted them into images using the \texttt{modelfit} task of {\sf Difmap}. 
Observation details and intensity maps are provided in Table~\ref{tab:sources} and Figure~\ref{fig:illustration}, respectively.

In each map, we covered the core (cf. Section~\ref{sec:mask}) with an elliptical mask about three times the size of the restoring beam in order to ensure that 
optically thick regions do not affect the correlation. After this, \vimap\ calculated the normalized 2D cross-correlation for the two input maps (cf. Equation~\ref{cor_formula}) and saved the resultant $r_{xy}$ array (Figure \ref{fig:result1}). The maximum correlation ($\approx$0.99) is found for a relative shift of $\approx$0.16 mas (cf. Table~\ref{tab:Shift_info}).

\begin{table}[t!]
\centering
\caption{Parameters of two VLBA data sets used for demonstration.}
\begin{tabular}{ccc}
\toprule
Frequency & Beam size$^{\rm a}$ & Noise$^{\rm b}$  \\ 
(GHz)     & (mas)               & (mJy/beam)	   \\
\midrule
  8.4 & $1.98 \times 0.92$ & 2.1 \\
 12.1 & $1.30 \times 0.64$ & 1.7 \\
\bottomrule
\end{tabular}
\tabnote{
{\sc Notes.}
Target: 3C 120.
Observing date: May 24, 2006.
Both maps have a size of 1024$\times$1024 pixels and use a pixel scale of 0.05 mas/pixel.
a: The 12-GHz map is convolved with the beam of the 8-GHz map before alignment.
b: r.m.s. noise estimated from source-free regions in the CLEAN map.}
\label{tab:sources}
\end{table}

\begin{table}[t!]
\centering
\caption{Shift of the 12-GHz map relative to the 8-GHz map.}
\begin{tabular}{lr}
\toprule
Shift in RA (pixels) & -3 \\
Shift in DEC (pixels) & 1 \\
2D shift (mas) & 0.158 \\
Maximum $r_{xy}$ & 0.992 \\
\bottomrule
\end{tabular}
\tabnote{}
\label{tab:Shift_info}
\end{table}

\begin{figure}[t!]
\centering
\includegraphics[angle=-90,width=82mm]{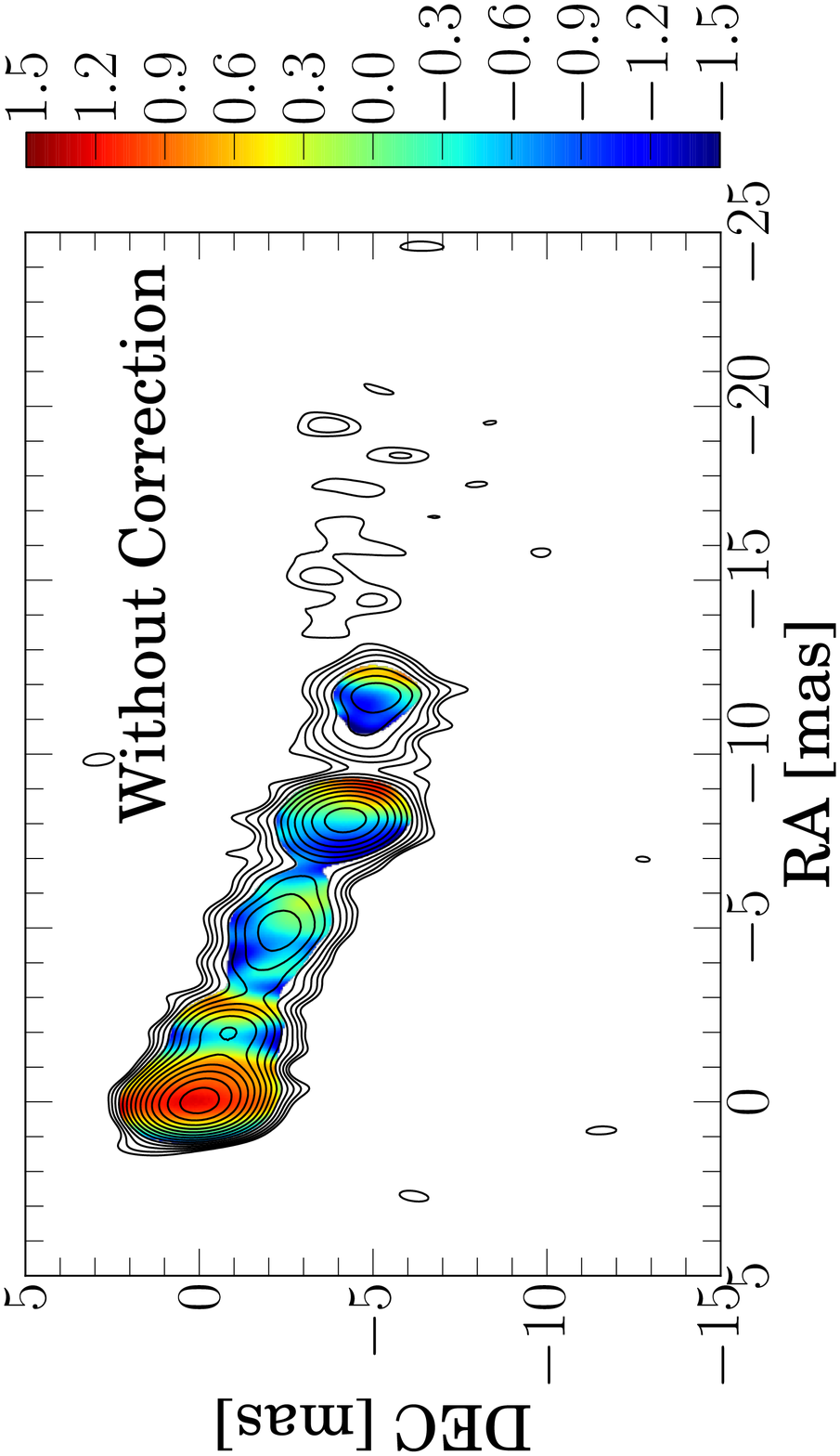} 
\includegraphics[angle=-90,width=82mm]{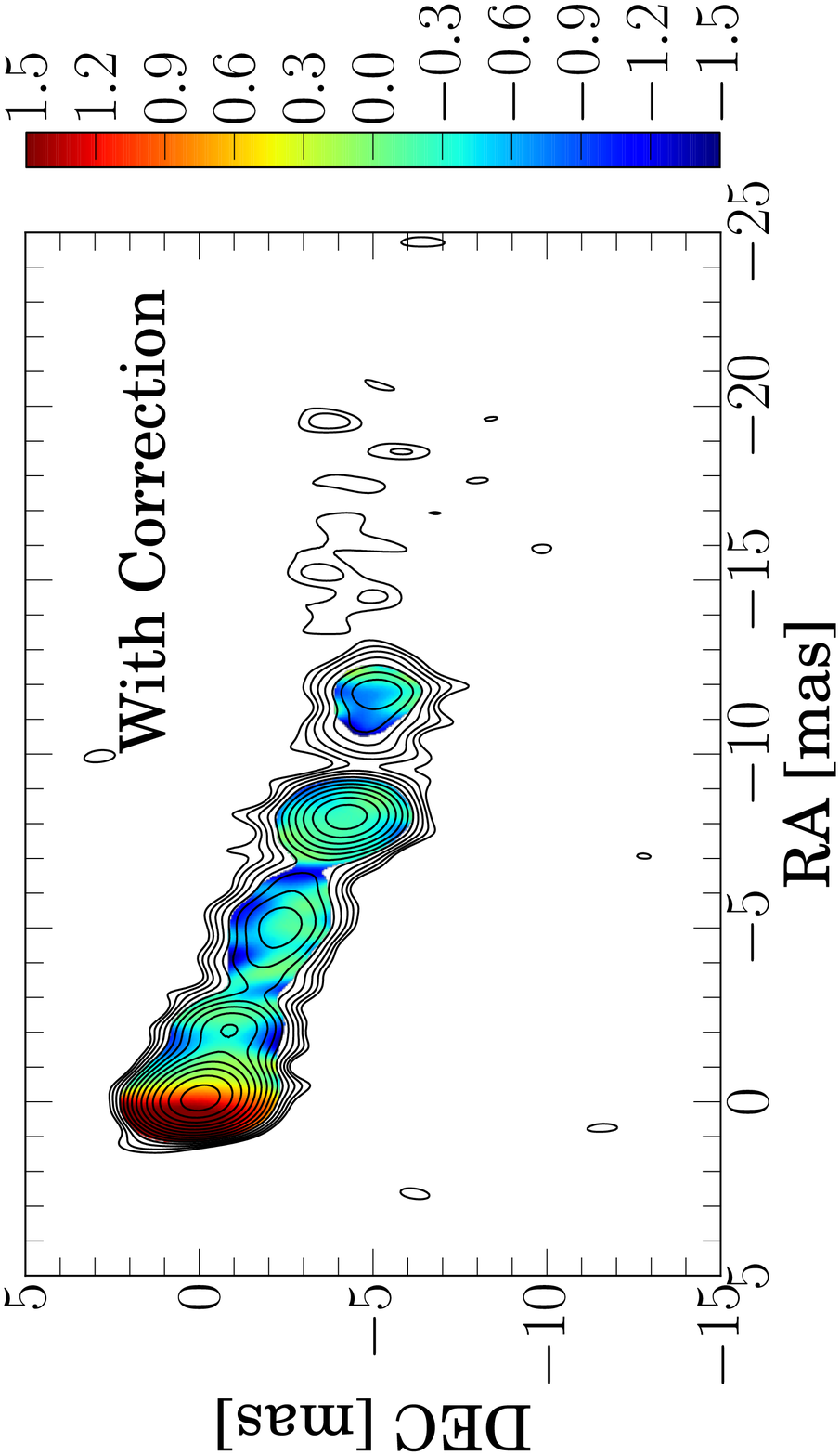} 
\caption{
Maps of spectral index $\alpha$ (color scale) of 3C 120 before \emph{(top)} and after \emph{(bottom)} correcting for the offset between the maps. The impact of the offset is evident.
}
\label{fig:result2_spix}
\end{figure}

\begin{figure}[t!]
\centering
\includegraphics[angle=-90,width=82mm]{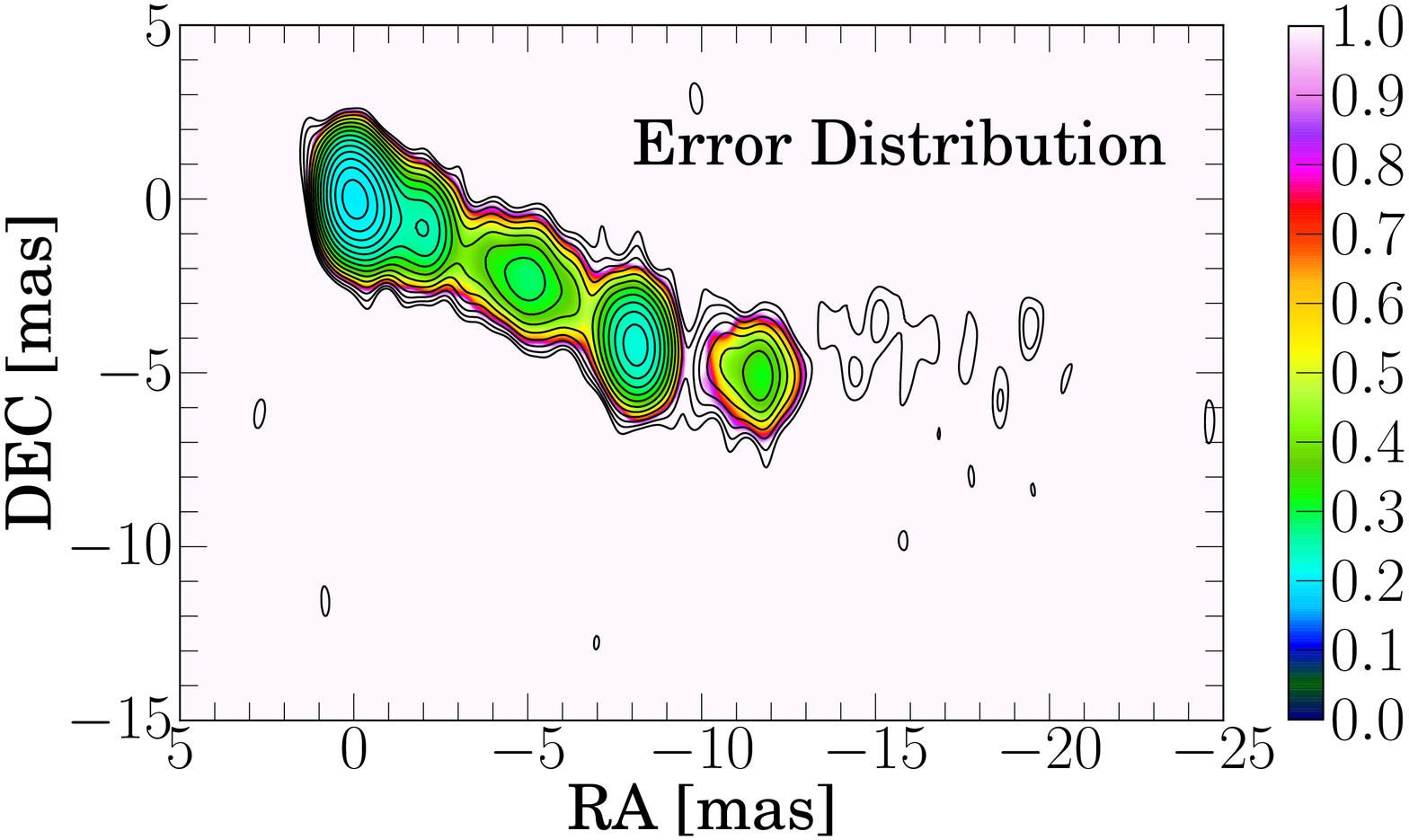} \\ 
\includegraphics[angle=-90,width=82mm]{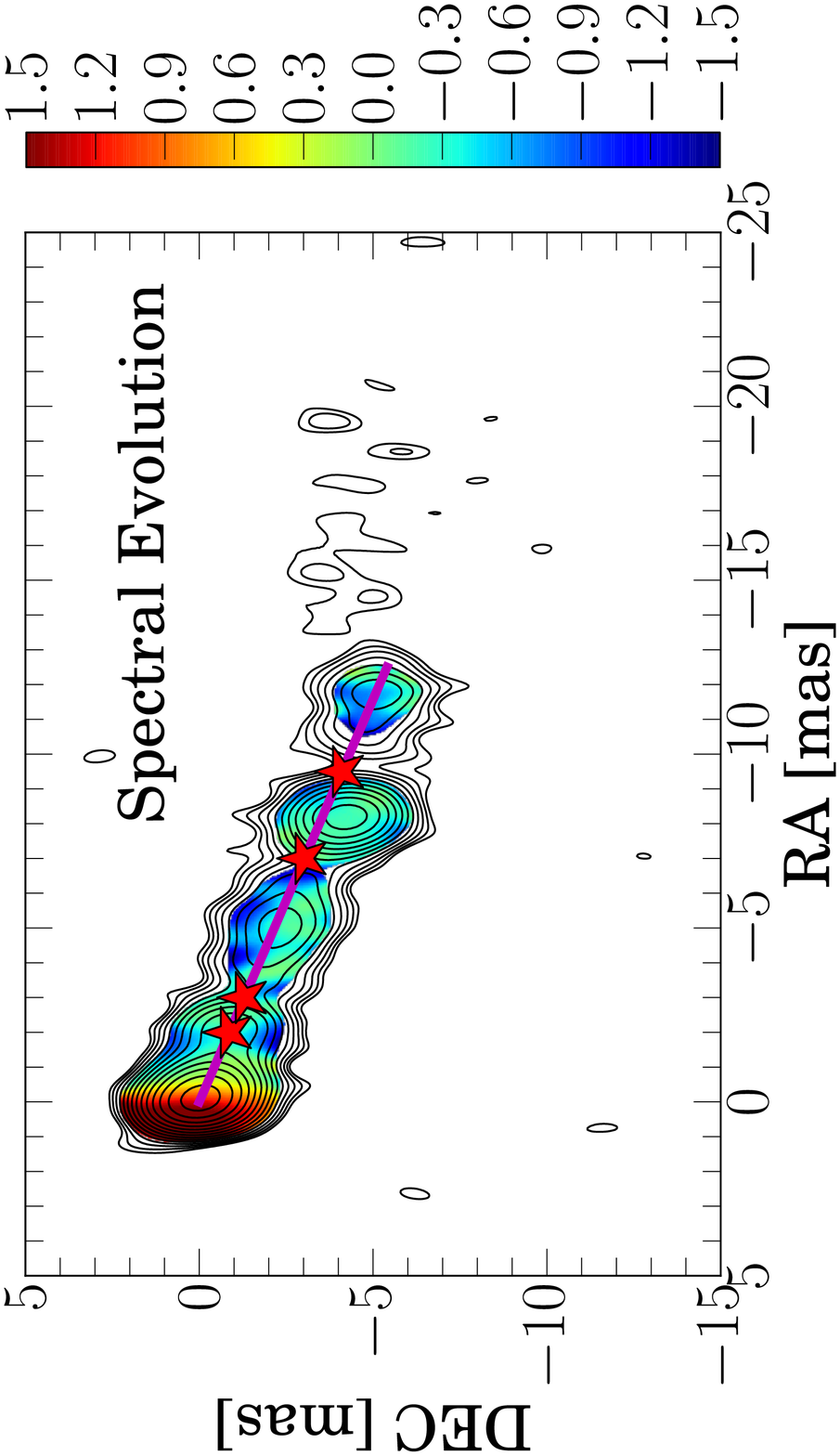} \\ 
\vskip2mm\hskip-3mm
\includegraphics[angle=-90,width=73mm]{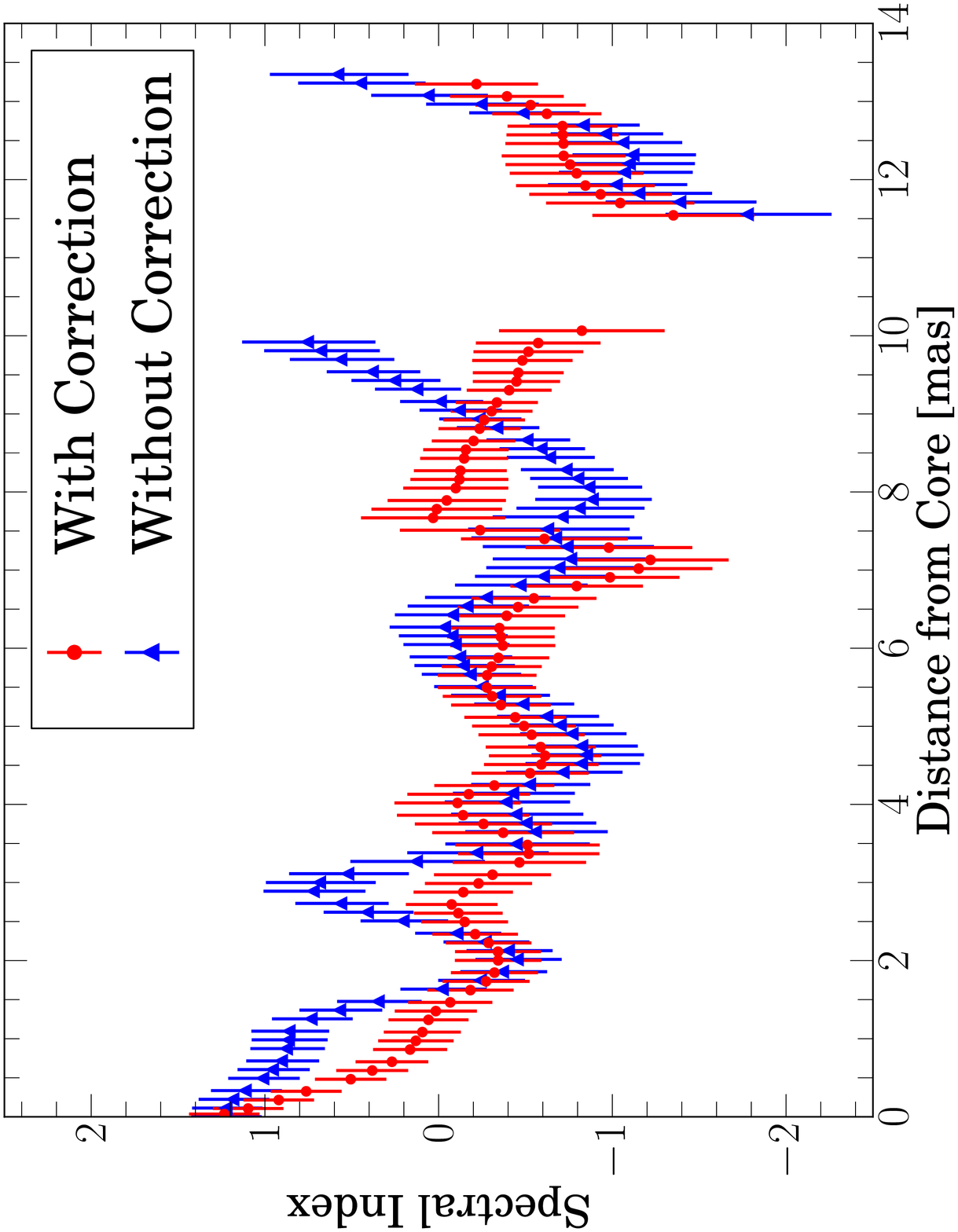} 
\caption{
\textit{Top}: Spectral index error map, assuming $\delta_{\nu}=0.1$ at both frequencies; the map was cut off at 0.5.
\textit{Center}: Same as the lower panel of Figure~\ref{fig:result2_spix}. The magenta line indicates a 1D cut through the $\alpha$ map along which the evolution of the index is analyzed. Red stars on the magenta line indicate locations of notable features shown in the bottom panel.
\textit{Bottom}: 1D spectral index profile along the cut indicated in the center panel, before and after correcting for the offset between the 8-GHz and 12-GHz maps.
}
\label{fig:result3_err_spixevol}
\end{figure}

After correcting for the relative offset, we computed the spectral index map (cf. Section~\ref{sec:indexmap}). We adopted an intensity limit of $0.9\%$ of the peak value of the 8-GHz map and an upper limit on the spectral index error of 0.5. Spectral index maps before and after image alignment are shown in Figure~\ref{fig:result2_spix}; the impact of the offset between the input maps is evident. Figure~\ref{fig:result3_err_spixevol} provides the spectral index error map (top panel) and the evolution of the spectral index along a one-dimensional cut following the jet (center and bottom panels).
At least three features should be noted:
(i) Up to 2 mas from the core ($\approx$ beam size; first red star from the core in the center panel of Figure \ref{fig:result3_err_spixevol}), the $\alpha$ profile obtained after image alignment rapidly decreases, whereas the profile obtained without alignment is rather flat.
(ii) About 3 mas from the core (second red star), the spectral index obtained without image alignment jumps up to $\approx 1$. Without image alignment, this feature might be incorrectly interpreted as an overdensity in the jet. (iii) About the same holds for the spectral index from 7 mas to around 10 mas (third to fourth red star).

\section{Summary\label{sec:summary}}

We report the development of a GUI-based interactive Python program, \vimap, for generating radio spectral index maps of active galactic nuclei from VLBI radio maps obtained at two different frequencies. For the crucial task of accurate spatial alignment of the input maps we employ the 2D cross-correlation algorithm of \citet{croke2008}.\footnote{
We note that \citet{croke2008} presented a C software program dedicated to the same purpose. However, their program comes with additional {\sf AIPS} dependencies which make it unsuitable for interactive data processing.} As yet, such a tool has not been available in the frame of standard radio interferometric data processing packages like {\sf AIPS} or {\sf CASA}. Our approach complements the method of image alignment via cross-identification of jet model components \citep{kameno2003,kadler2004}. As noted explicitly by \citet{kadler2004} (in their Section~4), the model component approach depends on a reliable cross-identification of optically thin jet components which is not always possible. \vimap\ offers a higher degree of flexibility because it does not rely on such cross-identifications and, in fact, does not even require modelling but may be applied to deconvolved (CLEANed) maps immediately. A detailed comparison of the two approaches is provided by \citet{fromm2013} (their Section~2.1).

A key feature of \vimap\ is its interactive GUI-based approach depending only on few standard, freely available, Python packages. Accordingly, \vimap\ provides users with a new tool capable of handling large amounts of multi-frequency VLBI data in a straightforward manner, especially in view of users of the Korean VLBI Network (KVN) \citep{lee2014} 
which is capable of \emph{simultaneous} multi-frequency observations, and the KVN and VERA Array (KaVA) \citep{niinuma2014}.

\acknowledgments

This work made use of data from the MOJAVE AGN monitoring database \citep{lister2009} and of the {\sf AIPS} and {\sf Difmap} software packages provided by the National Radio Astronomical Observatory of the USA. We acknowledge financial support from the Korean National Research Foundation (NRF) via Basic Research Grant 2012-R1A1A2041387.


\begin{thebibliography}{}


\bibitem[Allen et al.(2006)]{allen2006} Allen, S. W., Dunn, R. J. H., Fabian, A. C., et al. 2006, The Relation between Accretion Rate and Jet Power in X-ray Luminous Elliptical Galaxies, MNRAS, 372, 21
\bibitem[Blandford \& K\"{o}nigl(1979)]{bk79} Blandford, R. D., \& K\"{o}nigl, A. 1979, Relativistic Jets as Compact Radio Sources, ApJ, 232, 34
\bibitem[Boettcher et al.(2012)]{book2014} Boettcher, M., Harris, D. E., \& Krawczynski, H. 2012, Relativistic Jets from Active Galactic Nuclei (Weinheim: Wiley-VCH) 
\bibitem[Croke \& Gabuzda(2008)]{croke2008} Croke, S. M., \& Gabuzda, D. C. 2008, Aligning VLBI Images of Active Galactic Nuclei at Different Frequencies, MNRAS, 386, 619
\bibitem[Fromm et al.(2013)]{fromm2013} Fromm, C. M., Ros, E., Perucho, M., et al. 2013, Catching the Radio Flare in CTA 102. III. Core-Shift and Spectral Analysis, A\&A, 557, 105
\bibitem[Kadler et al.(2004)]{kadler2004} Kadler, M., Ros, E., Lobanov, A. P., et al. 2004, The Twin-Jet System in NGC 1052: VLBI-Scrutiny of the Obscuring Torus, A\&A, 426, 481
\bibitem[Kameno et al.(2003)]{kameno2003} Kameno, S., Inoue, M., Wajima, K., et al. 2003, Dense Plasma Torus in the GPS Galaxy NGC 1052, PASA, 20, 134
\bibitem[Kim \& Trippe(2013)]{kim2013} Kim, J.-Y., \& Trippe, S. 2013, How To Monitor AGN Intra-Day Variability at 230 GHz, JKAS, 46, 65
\bibitem[Lee et al.(2014)]{lee2014} Lee, S.-S., Petrov, L., Byun, D.-Y., et al. 2014, Early Science with the Korean VLBI Network: Evaluation of System Performance, AJ, 147, 77
\bibitem[Lister et al.(2009)]{lister2009} Lister, M. L., Aller, H. D., Aller, M. F., et al. 2009, MOJAVE: Monitoring of Jets in Active Galactic Nuclei with VLBA Experiments. V. Multi-Epoch VLBA Images, AJ, 137, 3718
\bibitem[Lobanov(1998)]{lobanov1998a} Lobanov, A. P. 1998, Ultracompact Jets in Active Galactic Nuclei, A\&A, 330, 79
\bibitem[Niinuma et al.(2014)]{niinuma2014} Niinuma, K., Lee, S.-S., Kino, M., et al. 2014, VLBI Observations of Bright AGN Jets with KVN and VERA Array (KaVA): Evaluation of Imaging Capability, PASJ, in press {\tt (arXiv:1406.4356)}
\bibitem[O'Sullivan \& Gabuzda(2009)]{osullivan2009} O'Sullivan, S. P., \& Gabuzda, D. C. 2009, Magnetic Field Strength and Spectral Distribution of Six Parsec-Scale Active Galactic Nuclei Jets, MNRAS, 400, 26
\bibitem[Park \& Trippe(2012)]{park2012} Park, J.-H., \& Trippe, S. 2012, Multiple Emission States in Active Galactic Nuclei, JKAS, 45, 147
\bibitem[Park \& Trippe(2014)]{park2014} Park, J.-H., \& Trippe, S. 2014, Radio Variability and Random Walk Noise Properties of Four Blazars, ApJ, 785, 76
\bibitem[Shepherd(1997)]{shep1997} Shepherd, M.-C. 1997, in ASP Conf. Ser. 125, Astronomical Data Analysis Software and Systems VI, ed. G. Hunt \& H. Payne (San Francisco: ASP), 77
\bibitem[Trippe(2014)]{trippe2014} Trippe, S. 2014, Does the Jet Production Efficiency of Radio Galaxies Control Their Optical AGN Types?, JKAS, 47, 159
\bibitem[Walker et al.(2000)]{walker2000} Walker, R. C., Dhawan, V., Romney, J. D., et al. 2000, VLBA Absorption Imaging of Ionized Gas Associated with the Accretion Disk in NGC 1275, ApJ, 530, 233
 

\end{thebibliography}
\end{document}